\RequirePackage{fix-cm}
\documentclass[twocolumn,epjc3]{svjour3}  

\usepackage{cite}
\usepackage{amsmath}
\usepackage{amssymb}
\usepackage{amsfonts}
\usepackage{appendix}
\usepackage{orcidlink}

\usepackage{lipsum}     
\usepackage{mathtools}
\usepackage{cuted}

\usepackage{graphics}
\usepackage{latexsym}
\usepackage{orcidlink}
\usepackage{enumerate}

\smartqed  
\RequirePackage{graphicx}
\journalname{Physics Letters B}
\begin{document}

	
\title{Anisotropic Dark Energy Stars within Vanishing Complexity Factor Formalism: Hydrostatic Equilibrium, Radial Oscillations, and Observational Implications
}


\author{
        Grigoris Panotopoulos \thanksref{e1,addr1}
        \orcidlink{0000-0002-7647-4072}
        \and
        \'Angel Rinc{\'o}n \orcidlink{0000-0001-8069-9162} \thanksref{e2,addr2}
        \and
        Il\'\i dio Lopes 
        \orcidlink{0000-0002-5011-9195}
        \thanksref{e3,addr3}
        \and
}


\thankstext{e1}{e-mail: 
\href{mailto:grigorios.panotopoulos@ufrontera.cl}
{\nolinkurl{grigorios.panotopoulos@ufrontera.cl}}
}

\thankstext{e2}{e-mail: 
\href{mailto:angel.rincon@ua.es}
{\nolinkurl{angel.rincon@ua.es}}
}

\thankstext{e3}{e-mail: 
\href{mailto:ilidio.lopes@tecnico.ulisboa.pt}
{\nolinkurl{ilidio.lopes@tecnico.ulisboa.pt}}
}


\institute{
Departamento de Ciencias F{\'i}sicas, Universidad de la Frontera, Casilla 54-D, 4811186 Temuco, Chile. \label{addr1}
           \and
Departamento de Física Aplicada, Universidad de Alicante, Campus de San Vicente del Raspeig, E-03690 Alicante, Spain. \label{addr2}
           \and
           Centro de Astrof{\'i}sica e Gravita{\c c}{\~a}o, Departamento de F{\'i}sica, Instituto Superior T{\'e}cnico-IST, Universidade de Lisboa-UL, Av. Rovisco Pais, 1049-001 Lisboa, Portugal. 
           \label{addr3}
}

\date{Received: date / Accepted: date}

\maketitle

\begin{abstract}
We investigate the structure and radial oscillations of anisotropic compact stars composed of dark energy, using the vanishing complexity factor formalism within general relativity.   This novel approach establishes a direct link between the energy density and anisotropic factor, providing a robust framework for studying these exotic stellar objects.  Employing an Extended Chaplygin Gas equation of state, we numerically compute interior solutions for both isotropic and anisotropic stars, revealing distinct differences in their properties. Additionally, we examine the oscillation modes and frequencies of these stars, highlighting the impact of anisotropy on their pulsational behavior. Our results reveal distinct differences in the stellar properties, such as the metric potentials, pressure, speed of sound, and relativistic adiabatic index, between the two cases.  Furthermore, we calculate the large frequency separation for the fundamental and first excited modes, offering insights relevant for future asteroseismology studies. Our findings shed light on the complex interplay of gravity, matter, and anisotropy in compact stars, providing a new perspective on dark energy's role in their structure and dynamics.
\keywords{
Relativistic stars \and Complexity factor \and Anisotropic stars \and Pulsating stars.}
\end{abstract}

\section{Introduction}\label{intro}

\smallskip\noindent 

The study of compact astrophysical objects, including black holes, neutron stars, and white dwarfs, has been central to relativistic and gravitational astrophysics since Karl Schwarzschild's groundbreaking 1916 solution to Einstein's field equations, which described the spacetime geometry around a non-rotating, spherically symmetric mass \cite{1916SPAW.......189S}.  This foundational work enhanced our understanding of black holes and opened avenues for exploring other compact objects' internal structures and properties, leading to numerous analytical and numerical solutions, first for isotropic stellar configurations and soon after for anisotropic ones.
 
 \smallskip\noindent 
 
Compact stars like neutron stars, which result from stellar evolution, represent a significant departure from conventional matter due to their immense internal densities, necessitating exploration through Einstein’s General Relativity (GR) \cite{Einstein:1915ca,Shapiro:1983du,Psaltis:2008bb,Lorimer:2008se,Zorotovic:2019uzl}.  Compact stars, including neutron stars, require a multidisciplinary approach involving nuclear particle physics, astrophysics, and gravitational physics. Neutron stars, the densest known celestial bodies after black holes, replicate conditions unachievable in Earth-based laboratories, serving as unique cosmic laboratories. Theoretical strange quark stars could represent the fundamental state of hadronic matter \cite{Alcock:1986hz,Alcock:1988re,Madsen:1998uh,Weber:2004kj,Yue:2006it,Leahy:2007we}.
 
 \smallskip\noindent 
 
Recent research has suggested the presence of strange matter in the cores of neutron star-hybrid stars \cite{Benic:2014jia,Yazdizadeh_2022,EslamPanah:2018rfe}, with some studies indicating similarities to conventional neutron stars \cite{Jaikumar:2005ne,2023ApJ...943...52R,2023PhRvD.107l3022R,2020PhRvD.101f3025S}. Strange quark stars might also explain super-luminous supernovae \cite{Ofek:2006vt,2009arXiv0911.5424O,2023EPJC...83.1065R}, with their potential existence across various astrophysical settings spurring further research \cite{Mukhopadhyay:2015xhs,Panotopoulos:2018ipq}.
 
 \smallskip\noindent 
 
Moreover, recent research has also shown a profound connection between dark matter and compact objects, as well as the pivotal role of dark energy in the universe's evolution \cite{2018RvMP...90d5002B,2008ARA&A..46..385F}. Dark matter, constituting approximately 27\% of the universe's mass-energy, remains elusive, its presence inferred solely through its gravitational influence on galaxies and cosmic structures \cite{2017FrPhy..12l1201Y}.  Conversely, dark energy, comprising roughly 68\% of the universe's energy density, is postulated to be the driving force behind its accelerated expansion.

\smallskip\noindent 

While the exact nature of dark matter and dark energy remains a mystery, several theories propose intriguing possibilities. Dark matter is hypothesized to consist of particles beyond the Standard Model, with leading candidates including Weakly Interacting Massive Particles (WIMPs), axions, and other exotic particles \cite{2010ARA&A..48..495F, 2015PhR...555....1B, 2018RPPh...81f6201R, 2019Univ....5..213P}. The nature of dark energy is even more enigmatic, but its potential connection to compact objects and their dynamics is an expanding area of research.

\smallskip\noindent 

Present and  upcoming space missions like Euclid \cite{2020A&A...642A.191E} and the Nancy Grace Roman Space Telescope \cite{2015arXiv150303757S}, alongside ground-based surveys like DESI \cite{2016arXiv161100036D}, are poised to provide unprecedented data that could revolutionize our understanding of these dark components. These cutting-edge experiments aim to unveil the properties of dark energy and could potentially illuminate the nature of dark matter, both of which are crucial for understanding the composition and evolution of the universe \cite{2016RPPh...79i6901W}.

\smallskip\noindent 

Several theoretical models propose that dark matter particles could interact with stars including compact objects, potentially leading to observable effects
\cite{2022panu.confE.313S,2023PhRvD.108h3028L,2024PhRvD.109d3043B,2023ApJ...958...49S,2023ApJ...953..115G,
2020PhRvD.102f3028I}. For instance, WIMPs could accumulate within neutron stars, eventually annihilating and heating the star's core, thereby influencing its thermal evolution \cite{2008PhRvD..77b3006K}. Axions, produced within main-sequence and neutron stars, could alter their structure and emission properties \cite{2016PhRvD..93f5044S,2023ApJ...943...95S,2021PhRvD.104b3008L}. Moreover, primordial black holes, a potential dark matter candidate, could be captured by neutron stars, resulting in mergers detectable through gravitational wave signals \cite{2013PhRvD..87l3524C}. These potential interactions underscore the importance of continued research and observation to validate these theoretical predictions and shed light on the elusive nature of dark matter.
 
 \smallskip\noindent 
 
 In all these classes of models mentioned above, anisotropic properties of matter play a unique role in the cosmos, influencing the behaviour and structure of astronomical objects like neutron stars. This characteristic is crucial for understanding phenomena such as dark matter, which might exhibit local anisotropy.  This is covered comprehensively in Ref. \cite{2022NewAR..9501662K}. In General Relativity, anisotropic fluids are well-studied, with their ability to follow geodesic paths highlighted in Ref.  \cite{2002JMP....43.4889H}. Research on their impact on the mass-radius relationship of anisotropic stars is significant \cite{2003RSPSA.459..393M}. 
 
 \smallskip\noindent 
 
Anisotropic stellar models offer a more realistic description of compact objects, as various physical phenomena like strong magnetic fields, superfluid states, or exotic matter can induce pressure anisotropy within the stellar interior. However, including anisotropy introduces complexity, requiring an extra degree of freedom to close the system of equations \cite{1997PhR...286...53H}.
 
 \smallskip\noindent 
 
 In this study, we focus on dark energy stars, which could offer insights into the roles of dark matter and dark energy in astrophysics and cosmology. These stars have been discussed in various scenarios, including as alternative explanations for observations attributed to compact stars and black holes, and their potential effects on neutron stars \cite{2019JCAP...07..012N,2021JCAP...02..045S,2012JCAP...03..037Y,2014arXiv1412.7323T}.
 
 \smallskip\noindent  
 
Here, we leverage the complexity factor formalism, rooted in the orthogonal decomposition of the Riemann tensor \cite{Herrera:2018bww}, to investigate anisotropic solutions to Einstein's field equations for compact stars. Focusing on the special case of a vanishing complexity factor, we establish a direct relationship between energy density and anisotropy, facilitating the exploration of these systems.

 \smallskip\noindent 
 
The momentum of this exploratory study is bolstered by the potential to detect gravitational waves from compact object mergers through observatories like LIGO \cite{2016PhRvL.116f1102A} and Virgo \cite{2005AIPC..794..307A}, and future projects like LISA \cite{2017arXiv170200786A} and the Einstein Telescope \cite{2010CQGra..27s4002P}. These observations provide unprecedented opportunities to probe matter under extreme conditions and test the limits of general relativity.

 \smallskip\noindent  
 
 Moreover, current X-ray observatories, including Chandra \cite{2009ApJ...697.1071A}, XMM-Newton \cite{2001A&A...365L...1J}, and NICER \cite{2014SPIE.9144E..20A}, have already provided valuable insights into neutron stars and other compact objects. Future missions like eXTP \cite{2019SCPMA..6229502Z}, STROBE-X \cite{2019arXiv190303035R}, Lynx \cite{2018arXiv180909642T}, and Athena promise to revolutionise our understanding of these extreme environments, particularly in studying stellar oscillations.
 
  \smallskip\noindent  
  
 We investigate anisotropic stars composed of dark energy, employing an Extended Chaplygin Gas equation of state. We compare our findings with isotropic models to discern the impact of anisotropy on stellar properties and radial oscillations. Additionally, we calculate the large separation of radial oscillations for these models, offering insights into their behaviour and opening a window for future observational measurements in neutron stars and other compact objects.
 
 \smallskip\noindent  
 
 Furthermore, we discuss a novel approach to studying anisotropic stars using the vanishing complexity factor concept derived from the Riemann tensor. This links energy density and anisotropy, providing a robust framework for investigating these stars and exploring their potential connection to dark matter.
 
 \smallskip\noindent 
 
The present paper is organised as follows: After this brief introduction, Section \ref{GR} defines the matter content and energy-momentum tensor for anisotropic stars. We derive the fundamental equations, including the anisotropic Tolman-Oppenheimer-Volkoff (TOV) equation and the mass conservation equation. Section \ref{VCF} introduces the complexity factor formalism to understand anisotropic stellar structures. In Section \ref{radial}, we examine the radial oscillations of anisotropic compact stars. We present the radial pulsation equations for isotropic stars and then discuss the modifications for anisotropic configurations. Section \ref{Dis} investigates anisotropic stars with a vanishing complexity factor, using the Extended Chaplygin Gas equation of state for dark energy. We numerically compute the interior solutions and compare them with isotropic counterparts. Throughout this manuscript, we use geometrical units where \( G = 1 = c \) and adopt the mostly negative metric signature in four space-time dimensions, \( \{+,-,-,-\} \).

\section{Anisotropic Stars: Matter Content, Metric Tensor and Structure Equations}\label{GR}
%
\smallskip\noindent 

In this section, we will define the matter content and the corresponding energy-momentum tensor for anisotropic stars. We will consider a spherically symmetric static object described using Schwarzschild-like coordinates.  The energy-momentum tensor and the metric potentials, which depend on the radial coordinate, will be used to derive the fundamental equations for the structure and equilibrium of these stars, including the anisotropic Tolman-Oppenheimer-Volkoff (TOV) equation and the mass conservation equation.

\subsection{Energy-Momentum Tensor and Metric tensor for Anisotropic Stars}
 
Let us start by defining the matter content to be used here. We will consider an energy-momentum tensor defined as follows:
\begin{align}
T^{\mu}_{\nu} \equiv \text{diag} \{ -\rho(r), P_{r}(r), P_{\perp}(r), P_{\perp}(r) \}.
\label{eq:Tmunu}
\end{align}
where, as always, the labels have the usual meaning: 
i) $\rho(r)$ is the energy density,  
ii) $P_r(r)$ is the radial pressure, 
iii) $P_{\perp}$  is the tangential pressure, and 
iv) \(\Pi(r) \equiv P_{\perp} - P_{r}\) represents the anisotropy factor, which is a useful auxiliary quantity. Please note that we have chosen to ignore the viscosity factor for simplicity. 

\smallskip\noindent 

In the previous  equation, matter contributes a distinct radial pressure \(P_{r}(r)\), whilst an anisotropic energy-momentum tensor results in a tangential pressure \(P_{\perp}(r)\). We represent the matter distribution using \(\Pi (r)=P_{\perp}(r) - P_{r}(r)\) to investigate interior solutions for relativistic stars, focusing on those with a strong component of anisotropic matter. Later, we shall discuss how \(\Pi (r)\) relates to the complexity factor formalism. This formalism provides a systematic approach to integrating anisotropies of various origins, for instance, those associated with ultra-dense relativistic objects
\cite{Sharif:2018pgq,Sharif:2018efi,Abbas:2018cha,Herrera:2019cbx}.

\smallskip\noindent

Next, we will assume a spherically symmetric static object, described using Schwarzschild-like coordinates: \(x^0 = t\), \(x^1 = r\), \(x^2 = \theta\), and \(x^3 = \phi\). The line element for the metric tensor is given by:
\begin{equation}
	\mathrm{d}s^2 = e^{\nu} \mathrm{d}t^2 - e^{\lambda} \mathrm{d}r^2 -
	r^2 \mathrm{d}\Omega^2.
	\label{metric}
\end{equation}
where $d\Omega^2\equiv \left( d\theta^2 + \sin^2\theta d\phi^2 \right)$ represents the element of solid angle and the metric potentials, $\nu(r)$ and $\lambda(r)$ depend only on the radial coordinate.

\subsection{Anisotropic Tolman-Oppenheimer-Volkoff Equations}

\smallskip\noindent 
By employing the Einstein field equations (disregarding the cosmological constant), the components of the energy-momentum tensor, expressed in local Minkowski coordinates and in terms of the metric potentials, can be written as:
\begin{eqnarray}
	\rho &=& -\frac{1}{8\pi}\left[-\frac{1}{r^2}+e^{-\lambda}
	\left(\frac{1}{r^2}-\frac{\lambda'}{r} \right)\right],
	\label{fieq00}
	\\
	P_r &=& -\frac{1}{8\pi}\left[\frac{1}{r^2} - e^{-\lambda}
	\left(\frac{1}{r^2}+\frac{\nu'}{r}\right)\right],
	\label{fieq11}
	\\
	P_\bot &=& \frac{1}{32\pi}e^{-\lambda}
	\left(2\nu''+\nu'^2 -
	\lambda'\nu' + 2\frac{\nu' - \lambda'}{r}\right),
	\label{fieq2233}
\end{eqnarray}
where prime symbols denote derivatives with respect to the radial coordinate $r$.

\smallskip\noindent 

At this point, we should obtain the anisotropic hydrostatic equilibrium equation, i.e., the well-known Tolman-Opphenheimer-Volkoff (TOV) equation for anisotropic stars.  Combining the previous equations for $\rho$,  $P_r $, and  $P_\bot $, we get:
\begin{equation} \label{Prp}
	P'_r = -\frac{1}{2}\nu'\left( \rho + P_r\right) + \frac{2}{r}\left(P_\bot-P_r\right).  
\end{equation}

\smallskip\noindent 
Even more, we can eliminate the $\nu'$-dependence in equation (\ref{Prp}) taking advantage of the following equation
\begin{equation}
	\frac{1}{2}\nu' =  \frac{m + 4 \pi P_r r^3}{r \left(r - 2m\right)}.
	\label{nuprii}
\end{equation}
Thus, substituting the equation  for $P'_r$, equation \eqref{nuprii} into equation \eqref{Prp}, we find the anisotropic TOV equation, namely
\begin{equation} \label{ntov}
	P'_r=-\frac{(m + 4 \pi P_r r^3)}{r \left(r - 2m\right)}\left( \rho + P_r\right)+\frac{2}{r}\left(P_\bot-P_r\right).
\end{equation}

\subsection{Mass Conservation Equation}

\smallskip\noindent 
The system of equations describing the structure and properties of a stellar object is made complete by including the differential equation governing the mass distribution within the star. This equation, known as the mass conservation equation, is given by:
\begin{align}
	m' &= 4 \pi r^2 \rho, \label{eq_mass}
\end{align}

\smallskip\noindent 
The mass conservation equation is derived from the Einstein field equations, which relate the geometry of spacetime to the matter content. In the context of a static, spherically symmetric spacetime, the relevant component of the Riemann curvature tensor, $R^3_{232}$, can be expressed as:
\begin{equation}
	R^3_{232}=1-e^{-\lambda}=\frac{2m}{r}.
	\label{rieman}
\end{equation}
 
%
\subsection{The Newton-like Interpretation of the TOV Equations}

It is worth noting that the Tolman-Oppenheimer-Volkoff (TOV) equation, which describes the equilibrium structure of a spherically symmetric star, can be broken down into terms that correspond to the various forces at play within the stellar interior. To gain a deeper understanding of these forces and their roles in maintaining the star's equilibrium, it is beneficial to identify and define three key components:

\begin{enumerate}
\item
The gravitational force, denoted as $F_g$, represents the attractive force that arises from the star's own gravity. This force acts to compress the stellar material, and its magnitude is determined by the mass distribution within the star. The expression for the gravitational force is given by:
\begin{equation}
	F_g = -\left( \frac{m + 4 \pi P_r r^3}{r \left(r - 2m\right)} \right)\left( \rho+ P_r\right),
\end{equation}
where $m$ is the mass enclosed within a radius $r$, $P_r$ is the radial pressure, and $\rho$ is the energy density.

\smallskip\noindent 

\item
The hydrostatic force, represented by $F_r$, arises from the pressure gradient within the star. This force acts to counterbalance the gravitational force, preventing the star from collapsing under its own weight. The hydrostatic force is defined as:
\begin{equation}
	F_r = -P'_r,
\end{equation}
where $P'_r$ denotes the radial derivative of the radial pressure.

\smallskip\noindent 

\item
The anisotropic force, symbolised by $F_p$, is a consequence of the pressure anisotropy that may exist within the stellar material. This force emerges when the tangential pressure, $P_\bot$, differs from the radial pressure, $P_r$. The anisotropic force is given by:
\begin{equation}
	F_p = \frac{2\Pi}{r},
\end{equation}
where $\Pi$ is the anisotropic factor, defined as the difference between the tangential and radial pressures: $\Pi = P_\bot - P_r.$
\end{enumerate}

\smallskip\noindent 

The presence of the anisotropic force in the TOV equation \eqref{ntov} can have significant implications for the star's structure and stability, depending on the nature of the pressure anisotropy. By expressing the TOV equation in terms of these three force components, we obtain a clear picture of the balance that must be maintained for the star to remain in equilibrium:
\begin{equation}
	F_g + F_r + F_p = 0.
\end{equation}

\smallskip\noindent 
This equation emphasizes the interplay between the gravitational, hydrostatic, and anisotropic forces in determining the star's structure and stability. The relative magnitudes and signs of these forces can lead to different scenarios:
\begin{enumerate}
\item
 When $F_p = 0$, the pressure is isotropic, and the TOV equation reduces to its standard form, describing the balance between the gravitational and hydrostatic forces.

\item
When $P_\bot > P_r$ (or $\Pi > 0$), the anisotropic force $F_p$ is positive and acts as a repulsive force, counteracting the combined attractive force of gravity and the hydrostatic force.

\item
 When $P_\bot < P_r$ (or $\Pi < 0$), the anisotropic force $F_p$ is negative and acts as an additional attractive force, working in conjunction with the gravitational force to compress the stellar material.
\end{enumerate}
By carefully analyzing the TOV equation and the roles of the gravitational, hydrostatic, and anisotropic forces, we can gain valuable insights into the complex interplay of factors that shape the structure and evolution of stars.
  
\subsection{Reformulated Energy-Momentum Tensor}

\smallskip\noindent

In the following, let us consider a reformulation of the energy-momentum tensor equation \eqref{eq:Tmunu} in a more convenient manner:
\begin{equation}
	T^{\mu}_{\nu}=\rho u^{\mu}u_{\nu}-  P
	h^{\mu}_{\nu}+\Pi ^{\mu}_{\nu},
	\label{24'}
\end{equation}
where we have introduced new parameters. Firstly, the four-velocity and four-acceleration are given by the expressions:
\begin{subequations}
	\begin{align}
		u^{\mu} &= (e^{-\frac{\nu}{2}},0,0,0),
		\\
		a^\alpha &= u^\alpha_{;\beta}u^\beta,
	\end{align}
\end{subequations}
where the non-vanishing component is $a_1 = -\nu^{\prime}/2$. Secondly,  the set $\{ \Pi^{\mu}_{\nu}, \Pi, h^{\mu}_{\nu}, s^{\mu}, P \}$ is defined as follows:
\begin{subequations}
	\begin{eqnarray}
		\Pi^{\mu}_{\nu} &=& \Pi\bigg(s^{\mu}s_{\nu}+\frac{1}{3}h^{\mu}_{\nu}\bigg) ,
		\\
		\Pi &=& P_{\bot}-P_r , \label{Delta}
		\\
		h^\mu_\nu &=& \delta^\mu_\nu-u^\mu u_\nu ,
		\\
		s^{\mu} &=& (0,e^{-\frac{\lambda}{2}},0,0),  \label{ese}
		\\
		P & \equiv & \frac{1}{3}\Bigl( P_{r}+2P_{\bot} \Bigl), 
	\end{eqnarray}
\end{subequations}
where the following properties hold:
$s^{\mu}u_{\mu}=0$,
$s^{\mu}s_{\mu}=-1$.
%
\subsection{Boundary Conditions for Matching Schwarzschild Spacetime}

We start solving the problem using the Schwarzschild spacetime, i.e.,
\begin{equation}
	\mathrm{d}s^2= \left(1-\frac{2M}{r}\right) \mathrm{d}t^2 - \left(1-\frac{2M}{r}\right)^{-1} \mathrm{d}r^2 -
	r^2  \mathrm{d}\Omega^2.
	\label{Vaidya}
\end{equation}
Thus, although it is possible to choose novel boundary conditions, the problem is closed by imposing appropriate boundary conditions on the surface $r=r_\Sigma \equiv  R$ for the Schwarzschild spacetime.
With the aid of the first and second fundamental forms across that surface, we have:
\begin{subequations}
	\begin{eqnarray}
		e^{\nu(r)} \Bigl|_{r=R} &=& 1-\frac{2M}{R},
		\label{enusigma}
		\\
		e^{\lambda(r)} \Bigl|_{r=R} &=& \Bigg(1-\frac{2M}{R}\Bigg)^{-1},
		\label{elambdasigma}
		\\
		P_r(r) \Bigl|_{r=R} &=& 0.
		\label{PQ}
	\end{eqnarray}
\end{subequations}
The last three equations represent the necessary and sufficient conditions for a smooth matching of the two metrics (\ref{metric}) and (\ref{Vaidya}) on the surface $r=R$. These boundary conditions ensure that the interior solution, described by the metric (\ref{metric}), seamlessly joins the exterior Schwarzschild spacetime (\ref{Vaidya}) at the stellar surface.

\section{The Complexity Factor: A Novel Approach to Anisotropic Stellar Structure}

\label{VCF}

\smallskip\noindent 

This section focuses on collecting the main (and also minimum) ingredients necessary to understand the concept of the complexity factor. In particular, we will consider the special case where the complexity factor is vanishing. Similarly, and given its relevance to this work, we mention how this approach can serve as a useful tool to make progress in the context of compact relativistic stars and some relativistic scenarios \cite{Herrera:2018bww}.

\smallskip\noindent 

Since Schwarzschild's 1916 solution, physicists have extensively studied models of stars based on Einstein's theory of relativity. This groundbreaking work led to various solutions describing different star structures, some of which also fulfill the necessary physical conditions within the star. This pioneering work opened a new window, and both exact and approximate solutions describing both isotropic and anisotropic stellar configurations have been the focus of thousands of physicists for decades.

\smallskip\noindent 

It is worth mentioning that in the case of anisotropic interior solutions, an additional condition is always needed to close the system, precisely the critical point that motivates us to move towards an alternative formalism or to introduce a well-developed ansatz to complete the set of equations and solve the corresponding system of differential equations. While there exist numerous potentially beneficial methods to close the system and obtain well-behaved solutions for compact stars, the concept of the complexity factor proves to be an invaluable tool. The strength of this approach lies in the fact that it is not an arbitrarily imposed additional constraint; instead, the formalism leverages the orthogonal decomposition of the Riemann tensor. Although the core idea itself is not entirely new, it was recently revisited in the seminal paper by Luis Herrera \cite{Herrera:2018bww}, which discussed the concept for self-gravitating systems in a static background. 

\smallskip\noindent 
This work presents a modern interpretation of complexity, motivated mainly by two shortcomings of previous descriptions:
\begin{enumerate}
\item
 Previous descriptions did not comprehensively include all components of the energy density fluid. They focused only on the energy density itself, neglecting possible additional contributions such as pressure.
\item
As shown in \cite{Sanudo:2008bu}, these earlier models considered the probability distribution with the energy density of the fluid distribution
\end{enumerate}

\smallskip\noindent 
Over the years, the concept underpinning the complexity factor formalism has been employed primarily from two complementary perspectives. The first approach is dedicated to investigating the formalism's impact on the exact solution of Einstein's field equations from a mathematical standpoint \cite{Sharif:2018pgq,Sharif:2018efi,Abbas:2018cha,Herrera:2019cbx}. This line of research focuses on the theoretical and analytical aspects of the formalism, exploring its mathematical properties and implications for solving the field equations.
However, the second approach proves to be more  interesting and practically relevant. Motivated by the notion of addressing or mitigating a weakness inherent in the study of stellar interiors, 
this approach allows us to go further in the study of compact objects since it circumvents the inclusion of some ansatze for the mass/density profile (as can be found in \cite{Panotopoulos:2018joc,Panotopoulos:2018ipq,Moraes:2021lhh,Gabbanelli:2018bhs,Panotopoulos:2019wsy,Lopes:2019psm,Panotopoulos:2019zxv,Abellan:2020jjl,Panotopoulos:2020zqa,Bhar:2020ukr,Panotopoulos:2020kgl,Panotopoulos:2021obe,Panotopoulos:2021dtu} and references therein). 

\smallskip\noindent 
 In this context, the second option clearly provides strong motivation to explore how the complexity factor formalism can complete the set of equations for stellar interiors. Indeed, by using this formalism, researchers can avoid ad hoc assumptions about mass or density distributions, resulting in more robust solutions and enabling the derivation of novel spherically symmetric solutions through the vanishing complexity factor approach.
Thus, the vanishing complexity formalism offers a novel approach to handling anisotropies in the context of compact stars, as exemplified by recent studies \cite{Arias:2022qrm,Andrade:2021flq,Rincon:2023zlp,Rincon:2023ens}.

\subsection{Orthogonal Decomposition of the Riemann Tensor and Scalar Functions}

\smallskip\noindent 

In the following, we will provide only the minimum information necessary to ensure that the article is self-contained. Please note that we have only highlighted a few key steps, and the detailed derivation can be found in \cite{Herrera:2018bww}. As we said, the key point is the orthogonal decomposition of the Riemann tensor for static self-gravitating fluids with spherical symmetry (see also \cite{Gomez-Lobo:2007mbg}).
Let us start by introducing three fundamental tensors:
  i) $Y_{\alpha \beta}$
 ii) $Z_{\alpha \beta}$
and
iii) $X_{\alpha \beta}$.
defined in terms of the Riemann tensor by
\begin{eqnarray}
Y_{\alpha \beta} &=& R_{\alpha \gamma \beta \delta}u^{\gamma}u^{\delta}, \label{electric} 
\\    
Z_{\alpha \beta} &=& R^{*}_{\alpha \gamma \beta
\delta}u^{\gamma}u^{\delta} = \frac{1}{2}\eta_{\alpha \gamma
\epsilon \mu} R^{\epsilon \mu}_{\quad \beta \delta} u^{\gamma}
u^{\delta}, \label{magnetic} 
\\
X_{\alpha \beta} &=& R^{*}_{\alpha \gamma \beta \delta}u^{\gamma}u^{\delta}=
\frac{1}{2}\eta_{\alpha \gamma}^{\quad \epsilon \mu} R^{*}_{\epsilon
\mu \beta \delta} u^{\gamma}
u^{\delta}. \label{magneticbis}
\end{eqnarray}
Here symbol $*$ is the dual tensor, so, 
\begin{equation}
   R^{*}_{\alpha \beta \gamma \delta}=\frac{1}{2}\eta_{\epsilon \mu \gamma \delta}R_{\alpha \beta}^{\quad \epsilon \mu} ,
\end{equation}
and $\eta_{\epsilon \mu \gamma \delta}$ represents the Levi-Civita tensor.
Then we have to link the tensors $Y_{\alpha \beta}, Z_{\alpha \beta}, X_{\alpha \beta}$ with the physical variables, which is done with the following relations
\begin{eqnarray}
Y_{\alpha\beta} &=& \frac{4\pi}{3}(\rho +3
P)h_{\alpha\beta}+4\pi \Pi_{\alpha\beta}+E_{\alpha\beta},\label{Y}
\\
Z_{\alpha\beta} &=& 0,\label{Z}
\\
X_{\alpha\beta} &=& \frac{8\pi}{3} \rho
h_{\alpha\beta}+4\pi
 \Pi_{\alpha\beta}-E_{\alpha\beta}.\label{X}
\end{eqnarray}
We have to define some supplementary quantities
$\{ E_{\alpha \beta} \equiv C_{\alpha \gamma \beta \delta}u^{\gamma}u^{\delta} , E \}$ which are defined via the following:
\begin{align}
E_{\alpha \beta} = E \bigg(s_\alpha s_\beta+\frac{1}{3}h_{\alpha \beta}\bigg),
\label{52bisx}
\end{align}
\begin{align}
E = -\frac{ e^{-\lambda}}{4}\left[ \nu ^{\prime \prime} + \frac{{\nu
^{\prime}}^2-\lambda ^{\prime} \nu ^{\prime}}{2} -  \frac{\nu
^{\prime}-\lambda ^{\prime}}{r}+\frac{2(1-e^{\lambda})}{r^2}\right].
\label{defE}
\end{align}
It should be noted that the tensor $E_{\alpha \beta}$ must fulfil the following properties:
 \begin{eqnarray}
 E^\alpha_{\,\,\alpha}=0,
 \quad 
 E_{\alpha\gamma} =  E_{(\alpha\gamma)},
 \quad 
 E_{\alpha\gamma}u^\gamma=0.
  \label{propE}
 \end{eqnarray} 

\smallskip\noindent 

At this point, it becomes essential to mention that the set $\{ Y_{\alpha\beta}, Z_{\alpha\beta}, X_{\alpha\beta} \}$ can be described directly in terms of extra scalar functions (see for example Ref. \cite{Herrera:2009zp}). Let us focus only on two of the last tensors, which are $X_{\alpha \beta}$ and $Y_{\alpha \beta}$ (in the static case), since they can be decoupled in terms of four scalars. 
With the help of the physical variables $\{ \rho, P_r, P_\bot \}$ we can write the new scalars accordingly:
\begin{align}
X_T     &= 8\pi  \rho ,  \label{esnIII} 
\\
X_{TF}  &= \frac{4\pi}{r^3} \int^r_0{\tilde r^3 \rho ' d\tilde r} \label{defXTFbis}, 
\\
Y_T     &= 4\pi( \rho  + 3 P_r-2\Pi) \label{esnV}, 
\\
Y_{TF}  &= 8\pi \Pi- \frac{4\pi}{r^3} \int^r_0{\tilde r^3 \rho' d\tilde r} \label{defYTFbis}.
\end{align}
Thus, the anisotropic factor $\Pi$ is rewritten using two of the previous scalar functions in the following form
\begin{equation}
	8\pi \Pi = X_{TF} + Y_{TF} \label{defanisxy}.
\end{equation}
Note that the complexity factor, $Y_{TF}$ in the previous set of equations, is inherently associated with the anisotropic factor.

\subsection{Anisotropic Stellar Structure and Vanishing Complexity Factor}

\smallskip\noindent

However, when the complexity factor $Y_{TF}$ vanishes identically, equation \eqref{defanisxy} establishes a direct relationship between the energy density and the anisotropic factor, providing a valuable link between these two crucial physical quantities. It is important to note that a concrete analytical form for the complexity factor is not yet well understood, and further research is needed to fully comprehend its mathematical structure and physical implications. As a first step towards making progress in this direction, we will impose the condition $Y_{TF}=0$, which leads to a simplified expression for the anisotropic factor $\Pi(r)$:
\begin{equation}
	\Pi(r) = \frac{1}{2r^3} \: \int^r_0{\tilde r^3 \rho'(\tilde r) d\tilde r} .
\end{equation}
This equation reveals that, under the assumption of a vanishing complexity factor, the anisotropic factor can be directly determined once a specific density profile $\rho(r)$ is prescribed for the self-gravitating system. It is crucial to emphasise two key points here:
i) The anisotropic factor for a self-gravitating system can be explicitly calculated once a particular density profile $\rho(r)$ is specified.
ii) The relationship between $\Pi(r)$ and $\rho'(r)$ gives rise to an integro-differential equation that, in general, must be solved numerically to obtain the behaviour of the anisotropic factor throughout the stellar interior.

\smallskip\noindent

By substituting the expression for $\Pi(r)$ into the anisotropic TOV equation, we obtain a modified form of the equation that governs the radial pressure gradient:
\begin{align}
P'_r=-\frac{(m + 4 \pi P_r r^3)}{r \left(r - 2m\right)}\left( \rho + P_r\right)+
	\frac{1}{r^4} \int^r_0{\tilde r^3 \rho'(\tilde r) d\tilde r}.
\end{align}
This equation can be equivalently expressed as a system of two coupled equations:
\begin{eqnarray}
	P'_r & = & -\frac{(m + 4 \pi P_r r^3)}{r \left(r - 2m\right)}\left( \rho + P_r\right)+\frac{2 \Pi}{r} 
\end{eqnarray}	
and
\begin{eqnarray}	
	\Pi' + \frac{3 \Pi}{r} & = & \frac{\rho'}{2},
\end{eqnarray}
subject to the initial condition $\Pi(r=0)=0$, which ensures regularity at the centre of the star.

\smallskip\noindent 

As a final remark, the application of the complexity factor formalism to realistic astrophysical scenarios remains an area great interest for further investigation. In this paper, we aim to address this gap by examining the implications of a vanishing complexity factor for relativistic compact stars, employing a generalised Chaplygin equation of state. We will then focus on the radial oscillations of these stars within this framework. Specifically, the following section will delve into a detailed study of the radial oscillations of these stars, utilising this novel approach.
By analysing anisotropic stars characterised by a vanishing complexity factor and contrasting our findings with those of isotropic stars, we aim to uncover the potential consequences, if any, of the assumption $Y_{TF}=0$.


\section{Radial Oscillations of Anisotropic Compact Stars}
\label{radial}

\smallskip\noindent 

In this section, we explore the radial oscillations of anisotropic compact stars, focusing on the modifications to the pulsation equations that arise due to the presence of pressure anisotropy. We begin by presenting the radial pulsation equations for isotropic compact stars and then proceed to discuss the additional terms that emerge in the case of anisotropic stellar configurations.

\subsection{Radial Pulsation Equations for Isotropic Compact Stars}

For a spherically symmetric system undergoing radial motion, the radial oscillation properties of a static equilibrium structure can be determined using Einstein's field equations, as demonstrated in \cite{1966ApJ...145..505B}. By analysing the radial displacement, $\Delta r$, and the pressure perturbation, $\Delta P_r$, we define the small perturbations of the dimensionless quantities $\xi = \Delta r/r$ and $\eta = \Delta P_r/P_r$, following the approach in Refs. \cite{1977ApJ...217..799C, Gondek:1997fd}. These perturbations are governed by the following equations:
\begin{equation}\label{ksi}
	\xi'(r) = -\frac{1}{r} \Biggl( 3\xi + \frac{\eta}{\Gamma} \Biggr) - \frac{P_r'}{\zeta} \xi,
\end{equation}
\begin{equation}\label{eta}
	\begin{split}
		\eta'(r) = \xi \Biggl[ \omega^{2} r \frac{\zeta}{P} e^{\lambda - \nu } -\frac{4P_r'}{P_r} -8\pi \zeta re^{\lambda} \\
		+  \frac{r(P_r')^{2}}{P_r \: \zeta} \Biggr]
		 + \eta \Biggl[ -\frac{\rho P'}{P_r \zeta} -4\pi \zeta re^{\lambda}\Biggr] ,
	\end{split}
\end{equation}
where we have defined a new function $\zeta \equiv P_r+\rho$,
 $\omega$ represents the frequency oscillation mode, and $\Gamma$ denotes the relativistic adiabatic index, given by
\begin{equation}
	\Gamma = \Biggl(1+\frac{\rho}{P_r} \Biggr) c_s^{2},
\end{equation}
with $c_s^{2}$ being the square of the speed of sound, expressed as
\begin{equation}\label{cs}
	c_s^{2} = \Biggl(\frac{dP_r}{d\rho}\Biggr).
\end{equation}
The two coupled differential equations, Eqs.~\eqref{ksi} and \eqref{eta}, are supplemented with two boundary conditions: one at the centre, where $r=0$, and another at the surface, where $r$ = $R$. The boundary condition at the centre requires that
\begin{equation}
	\eta = -3\Gamma \xi 
\end{equation}
be satisfied. Furthermore, Eq.~\eqref{eta} must be finite at the surface, implying that
\begin{equation}
	\eta = \xi \Biggl[ -4 +\left(1-2\frac{M}{R}\right)^{-1} \Biggl( -\frac{M}{R} -\frac{\omega^{2} R^{3}}{M}\Biggr)\Biggr]
\end{equation}
must hold, where $M$ and $R$ correspond to the mass and radius of the star, respectively. The frequencies are computed using
\begin{equation}
	\nu = \frac{\omega}{2\pi} = \frac{s \: \omega_0}{2\pi} ~~(\text{kHz}),
\end{equation}
where $s$ is a dimensionless number, and $\omega_0 \equiv \sqrt{M/R^3}$.

\smallskip\noindent 

We use the shooting method analysis, where one starts the integration for a trial value of $\omega^2$ and a given set of initial values that satisfy the boundary condition at the center. We integrate towards the surface, and the discrete values of $\omega^2$ for which the boundary conditions are satisfied correspond to the eigenfrequencies of the
radial perturbations.

\smallskip\noindent 

These equations represent the Sturm-Liouville eigenvalue equations for $\omega$. The solutions provide the discrete eigenvalues $\omega_n^{2}$ and can be ordered as 
$ \omega_0 ^{2} < \omega_1 ^{2} <... <\omega_n ^{2}, 
$
where $n$ is the number of nodes for a star of a given mass and radius. Finally, once the spectrum is known, the so called large frequency separation may be computed
\begin{equation}
	\Delta \nu_n = \nu_{n+1} - \nu_n, \; \; \; n=0,1,2,3,...
\end{equation}
or in other words the difference between consecutive modes, which is widely used in Asteroseismology to learn about star properties, inner structure and composition.


\subsection{Modifications to the Radial Pulsation Equations for Anisotropic Stars}

The above equations for radial perturbations enable us to study the radial oscillation modes of stars composed of isotropic matter. However, in the case of anisotropic stars, there are additional terms that are proportional to the anisotropic factor. Consequently, the linear system of coupled perturbations takes the following form \cite{Arbanil:2021ahh}:
\begin{equation}\label{ksi1}
	\xi' (r) = -\frac{1}{r} \Biggl( 3\xi + \frac{\eta}{\Gamma} \Biggr) - \left( \frac{P_r'}{\zeta} + \frac{2 \Pi}{r P_r \Gamma} \right) \xi,
\end{equation}
\begin{equation}\label{eta1}
\begin{split}
\eta'(r) = \xi \Biggl[ \omega^{2} r \frac{\zeta}{P_r}
 e^{\lambda - \nu } - \frac{4P_r'}{P_r} -8\pi \zeta re^{\lambda} \frac{P_r+\Pi}{P_r} \\
+  \frac{8 \Pi}{r P} + \frac{r(P_r')^{2}}{P_r \: \zeta} \Biggr] + \eta \Biggl[ -\frac{\rho P_r'}{P_r \: \zeta} -4\pi \zeta re^{\lambda}\Biggr] ,
\end{split}
\end{equation}
The presence of the anisotropic factor $\Pi$ introduces modifications to the previous equations governing the radial oscillations of anisotropic stars. These additional terms account for the effects of pressure anisotropy on the oscillatory behaviour of the stellar configuration. By solving these coupled equations, subject to the appropriate boundary conditions, one can investigate the radial oscillation modes and the corresponding eigenfrequencies for anisotropic stars, thereby gaining valuable insights into their stability properties and the impact of anisotropy on their pulsational characteristics.

\section{Anisotropic Compact Stars with Vanishing Complexity Factor: A Study with the Extended Chaplygin Equation of State}\label{Dis}

\smallskip\noindent

In the present work we have investigated anisotropic stars made of dark energy within the framework of General Relativity, taking advantage of the vanishing complexity formalism. We have numerically computed interior solutions of realistic spherical configurations of anisotropic matter and compared our solutions against isotropic ones (i.e., with $\Pi=0$). 
Following  Ref. \cite{DEstars2}, we have considered an Extended Chaplygin Gas equation-of-state for dark energy (see also Refs. \cite{Pourhassan:2014ika,Pourhassan:2014cea}):
\begin{equation}
	\displaystyle P_r(r) = - \frac{B^2}{\rho (r)} + A^2 \rho (r).
\end{equation}

\smallskip\noindent

where the two parameters $\{A, B\}$ are positive constants, with $A$ being dimensionless and $B$ being a parameter with dimensions of energy density and pressure. Furthermore, it is worth remembering that the standard Chaplygin's equation-of-state is given by $P_r = -B^2/\rho$ \cite{chaplygin1}, while the additional barotropic term corresponds to $A^2\rho$, which also leads to an excellent dark energy model \cite{Panotopoulos:2020qbx}. 
The numerical values of $A$ and $B$ considered here are as follows:
\begin{equation}
	A = \sqrt{0.45}, \; \; \; \; B = 0.2 \times 10^{-3}~km^{-2}.
\end{equation}
From the specific form of the equation-of-state, it is evident that the pressure and density cannot simultaneously become zero, as sometimes occurs. Thus, a vanishing pressure at the surface of the star is equivalent to a non-vanishing value for the energy density, which is found to be $\rho_s = B/A$.

\subsection{Background Solutions: Isotropic vs. Anisotropic Stars with Identical Mass and Radius}

\begin{figure*}[ht!]
	\centering
	\includegraphics[scale=0.935]{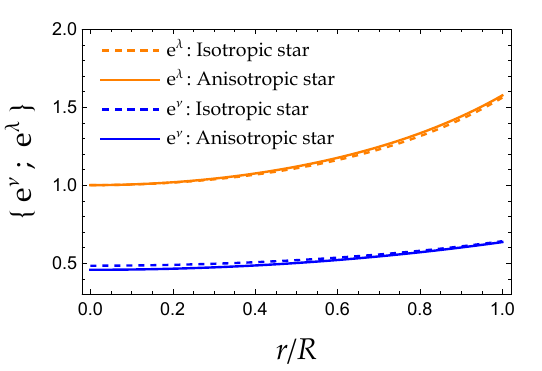} \
	\includegraphics[scale=0.935]{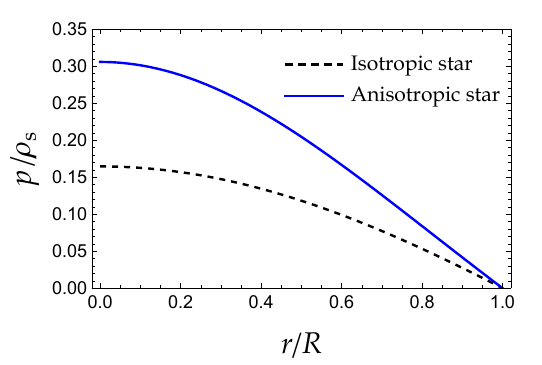} \caption{
		Metric potentials $\{ \text{e}^{\nu} ; \text{e}^{\lambda} \}$ and normalized pressure versus normalized radial coordinate for anisotropic (dashed curves) and anisotropic (solid lines) stars.
		{\bf{Left Panel:}} Metric potentials against the normalized radial coordinate.
		{\bf{Right Panel:}} Normalized pressure against the normalized radial coordinate.
	}
	\label{fig:1} 	
\end{figure*}

\smallskip\noindent
Firstly, regarding background solutions, we have numerically integrated the TOV equations, imposing the appropriate initial conditions at the centre of the stars, $r=0$. This task has been performed twice: for anisotropic objects within the vanishing complexity factor formalism and for stars made of isotropic matter, considering the same stellar mass and radius, $M=1.4\; M_{\odot}$, $R=11.6\;{\rm km}$, for comparison purposes. The quantities of interest, namely metric potentials, pressure, speed of sound, and relativistic adiabatic index, are displayed in Fig. \ref{fig:1} and \ref{fig:2} as functions of the normalized radial coordinate, $r/R$, both for isotropic stars (dashed lines) and anisotropic objects (solid lines). The pressure remains finite at the origin and monotonically decreases throughout the stars until it becomes zero at the surface. Furthermore, causality is not violated, as the sound speeds take values in the range $0 < c_s < 1$ from the centre to the surface of the objects. It is noteworthy that the speed of sound of isotropic stars is higher than that of their anisotropic counterparts. The same holds for the relativistic adiabatic index, whereas the pressure of isotropic stars is lower than that of stars made of anisotropic matter.

\begin{table}
	\caption{Frequencies (in kHz) of radial oscillation modes both for isotropic and anisotropic stars considering $M=1.4 M_{\odot}$ and $R=11.6 km$.}
	\label{tab:1}       
	\begin{tabular}{lll}
		\hline\noalign{\smallskip}
		Mode order $n$  & Anisotropic star & Isotropic star  \\
		\noalign{\smallskip}\hline\noalign{\smallskip}
		0  &  6.19 &  6.73 \\ 
		1  & 14.50 & 15.23 \\ 
		2  & 22.29 & 23.30 \\
		3  & 29.97 & 31.28 \\
		4  & 37.60 & 39.22 \\ 
		5  & 45.22 & 47.14 \\ 
		6  & 52.82 & 55.06 \\ 
		7  & 60.41 & 62.96 \\
		8  & 68.00 & 70.87 \\ 
		9  & 75.59 & 78.77 \\ 
		10 & 83.17 & 86.66 \\ 
		\noalign{\smallskip}\hline
	\end{tabular}
\end{table}

\subsection{Radial Oscillations: Numerical Results for Isotropic and Anisotropic Stars}

\begin{figure*}[ht!]
	\centering
	\includegraphics[scale=0.935]{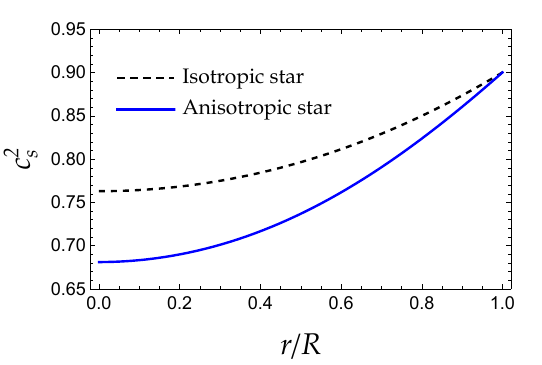} \
	\includegraphics[scale=0.935]{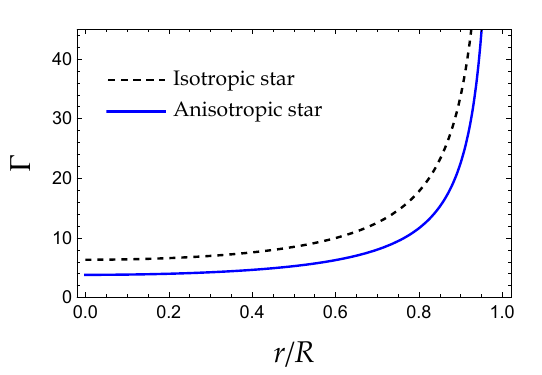} 
	\caption{
		Speed of sound $c_s^2$ and adiabatic relativistic index  $\Gamma$ versus normalized radial coordinate for isotropic (dashed curves) and anisotropic (solid lines) stars.
		{\bf{Left Panel:}} Speed of sound against the normalized radial coordinate.
		{\bf{Right Panel:}} Adiabatic relativistic index against the normalized radial coordinate.
	}
	\label{fig:2} 	
\end{figure*}

\begin{figure*}[ht!]
	\centering
	\includegraphics[width=0.49\textwidth]{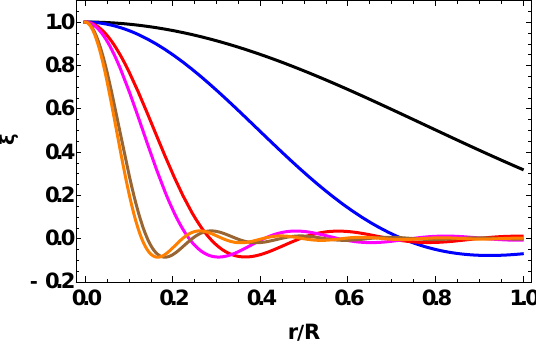} \
	\includegraphics[width=0.49\textwidth]{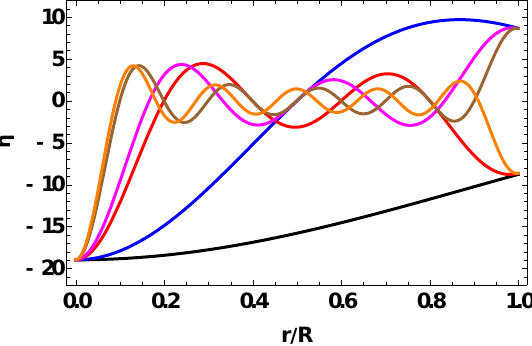}
	\caption{
		Radial profiles of the perturbations. 
		{\bf{Left Panel:}} The eigenfunctions $\xi(r)$ versus $r/R$ for the fundamental mode as well as the excited modes $n=1,4,5,9,10$.
		{\bf{Right Panel:}} $\eta(r)$ versus $r/R$ for the fundamental mode as well as the excited modes $n=1,4,5,9,10$.
	}
	\label{fig:3} 	
\end{figure*}

\begin{figure*}[ht!]
	\centering
	\includegraphics[width=0.49\textwidth]{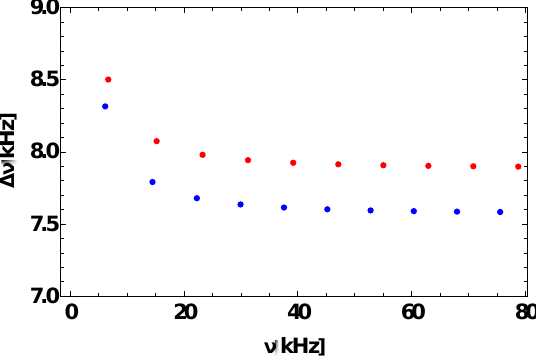} \
	\includegraphics[width=0.49\textwidth]{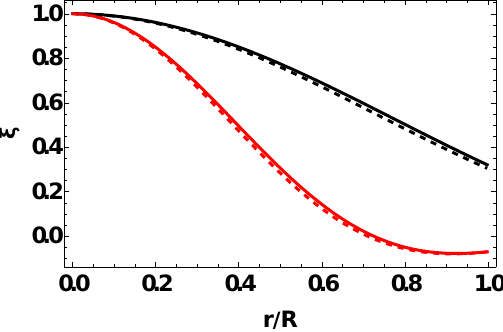}
	\caption{
		{\bf{Left Panel:}} Large frequency separations for isotropic (red points) and anisotropic stars (blue points). {\bf{Right Panel:}} Fundamental and first excited modes for isotropic (dashed curves) and anisotropic (solid lines) stars (right panel).
	}
	\label{fig:4} 	
\end{figure*}

\smallskip\noindent
Next, concerning the radial oscillations of pulsating stars, we have numerically integrated the linearised system of coupled perturbations, imposing the appropriate boundary conditions both at the centre and at the surface of the stars. 
The condition at $r=R$ is satisfied only for certain allowed values of the frequencies, which are displayed in Table \ref{tab:1}. Our results show that the frequencies of the objects made of isotropic matter are larger than those of their anisotropic counterparts. The corresponding eigenfunctions, as well as the large frequency separations, have also been computed and are shown in Fig. \ref{fig:3} and the left panel of Fig. \ref{fig:4}. Fig. \ref{fig:3} displays the radial profiles of both perturbations of isotropic stars for the lowest, intermediate, and highly excited modes, while the radial profiles of anisotropic stars are qualitatively very similar, the main difference being the value of the $\eta$ function at the origin, due to the fact that the relativistic adiabatic index is somewhat different, as seen in the right panel of Fig. \ref{fig:2}. We observe that the radial profiles of the perturbations exhibit the typical behaviour observed in any Sturm-Liouville problem, namely, the number of zeros of the eigenfunctions equals the value of the overtone number $n=0,1,2,...$. Furthermore, we note that the large frequency separations tend to a certain constant for highly excited modes, in accordance with the asymptotic theory. The asymptotic value for isotropic stars is higher than that of the objects made of anisotropic matter. Finally, in the right panel of Fig. \ref{fig:4}, we have made a comparison between isotropic and anisotropic stars regarding the fundamental, $n=0$, and the first excited modes, $n=1$. We observe that the radial profiles of the stars made of isotropic matter (dashed lines) remain slightly below the radial profiles of the anisotropic stars (solid lines).


\section{Conclusion}\label{Conc}

\smallskip\noindent

To summarize our work, in the present article we have studied radial oscillation modes of pulsating stars made of anisotropic matter within the vanishing complexity factor formalism. First we integrated numerically the structure equations imposing the usual initial conditions at the center of the star in order to compute the stellar mass and radius as well as the quantities of interest, such as pressure, metric potentials etc. Next, we integrated numerically the equations for the perturbations imposing the appropriate boundary conditions both at the center and at the surface of the star in order to compute the spectrum as well as the corresponding eigenfunctions. The large frequency separations have been computed as well. Moreover, we have made a comparison between the anisotropic stars and their isotropic counterparts considering the same stellar mass and radius. Our main numerical results are summarized in one table and several figures. According to our findings, isotropic objects are characterized by larger frequencies and also by a higher speed of sound throughout the stars.

\smallskip\noindent

Finally, we reiterate that radial oscillations in relativistic stars are crucial in astrophysics because of their implications for understanding stellar stability, the equation of state (EoS) of ultradense matter, and the dynamics of compact objects. These oscillations serve as essential diagnostic tools for probing the internal structure of neutron stars, providing insight into the behavior of matter at extreme densities and pressures \cite{Chandrasekhar1964, Lindblom1984}. The study of the frequencies and damping rates of radial oscillations allows the identification of conditions leading to stellar collapse into black holes or supernova explosions \cite{Shapiro1983}.

\smallskip\noindent

In conclusion, the investigation of radial oscillations not only enhances our understanding of neutron star dynamics and stability but also serves as a bridge between theoretical predictions and observational data, linking macroscopic astrophysical phenomena with microscopic physical laws. By studying such oscillations, we gain valuable insights into the fundamental properties of matter under extreme conditions and further our comprehension of the complex interplay between gravity and matter in the context of compact astrophysical objects.

\section*{Acknowledgments}

A.~R. acknowledges financial support from the Generalitat Valenciana through PROMETEO PROJECT CIPROM/2022/13.
A.~R. is funded by the Mar{\'i}a Zambrano contract ZAMBRANO 21-25 (Spain) (with funding from NextGenerationEU).
I.~L. thanks the Funda\c c\~ao para a Ci\^encia e Tecnologia (FCT), Portugal, for the financial support to the Center for Astrophysics and Gravitation (CENTRA/IST/ULisboa)  through the Grant Project No.~UIDB/00099/2020  and Grant No.~PTDC/FIS-AST/28920/2017. 



%

\bibliographystyle{unsrt}
\bibliography{biblioIL2.bib}

%

\end{document}